\documentstyle[12pt]{article}
\begin{document}
\renewcommand{\theequation}{\thesection.\arabic{equation}}
\thispagestyle{empty} \vspace*{-1.5cm} \hfill {\small SU-ITP}
\\[8mm]

\setlength{\topmargin}{-1.5cm} \setlength{\textheight}{22cm}
\begin{center}
{\large Holography in the Flat Space Limit }\\

\vspace{2 cm} {\large Leonard Susskind}\\ Department of Physics,
Stanford University\\ Stanford, CA 94305-4060 \\ \vspace{5cm}
\begin{abstract}
Matrix theory and the AdS/CFT correspondence provide
nonperturbative holographic formulations of string theory. In both
cases the finite N theories can be thought of as infrared
regulated versions of flat space string theory in which removing
the cutoff is equivalent to letting N go to infinity.

In this paper we consider the nature of this limit. In both cases
the holographic mapping  becomes completely nonlocal. In matrix
theory this corresponds to the growth of D0-brane bound states
with N. For the AdS/CFT correspondence there is a similar
delocalization of the holographic image of a system as N
increases.   In this  case the limiting theory seems to require a
number of degrees of freedom comparable to  large N matrix quantum
mechanics.
\end{abstract}
\vspace{1cm} {\it December 1998}
\end{center}
\newpage

\setcounter{equation}{0}
\section{Introduction}

According to the holographic principle,  a physical system  of
dimensionality D which includes gravity,  should be described by a
quantum system which lives in fewer dimensions. We have seen a
good deal of evidence for the holographic principle from both
matrix theory and the AdS/CFT correspondence but very little real
understanding of how it works, in other words, how a general
configuration of a D dimensional system is coded by lower
dimensional degrees of freedom. My main purpose in this paper is
to provoke discussion about the mechanism of holography \cite{1},
\cite{2}. Most of the things that I will discuss I do not
understand very well. In trying to formulate them precisely I have
mainly encountered frustration. Nevertheless I think they are
important and deserve to be discussed.

One of the characteristic features of a real hologram is that it
codes information in a highly nonlocal way. For example by
casually looking at a hologram of several distinct objects it is
impossible to tell how many objects it describes or their size and
shape. These details are completely delocalized on the hologram.
The point of this paper is to argue that quantum gravity is
holographic in exactly this sense.

Two concrete realizations of holographic theories now exit, namely
matrix theory \cite{3} and the AdS/CFT correspondence
\cite{4},\cite{5},\cite{6},\cite{7}. In both theories the hologram
is a large $N$ super Yang Mills (SYM) theory. Furthermore in both
cases $N$ serves as a kind of infrared regulator. In the limit $N
\to \infty$ keeping the Yang Mills coupling fixed both theories
describe physics in infinite flat space. Furthermore, as we shall
see, as $N$ grows, the mapping between the hologram and the system
it describes becomes more and more nonlocal, In this respect the
mapping is like a real hologram. In this paper I will raise some
unanswered questions about the nature of the holographic mapping,
especially in the limit of infinite flat spacetime. As we shall
see, the large $N$ limit involved in going to flat space is quite
different  than the usual 't Hooft limit in which the coupling
shrinks to zero as $N$ increases. The flat space limits in Matrix
and   AdS/CFT theories both involve letting $N$ go to infinity
with fixed gauge coupling. Thus the 't Hooft coupling parameter
$g_{ym}^2 N$ tends to infinity and the fixed point becomes
infinitely strongly coupled.

Imagine a system composed of point sources of light (particles).
Assume that the  light from the different sources is coherent as
long as they are within a coherence length $L_c$. All of this
takes place in the 3-dimensional half space $z>0$. At $z=0$ in the
$x,y$ plane there is a photographic film which records the light
from the particles. As long as the particles are separated by
distance greater than $L_c$ they form two separate blobs of light
on the film. If we made a movie from such photos we   could follow
the individual particles' motion from these blobs. However as soon
as they approached within $L_c$ the individual identities would
disappear. However the details would not be lost. At this point
the details such as the number and position of point sources would
become  encoded holographically, that is nonlocally distributed
over the coherence length $L_c$.  As the coherence length
increases the information becomes completely delocalized over the
entire hologram.  For an ordinary hologram the information is in
the interference patterns created by the  coherent light sources.
For matrix theory and the AdS/CFT correspondence the coding is
more obscure but in both cases it involves the $N\times N$ matrix
degrees of freedom of Super Yang Mills theories. In both these
theories we will see the same kind of delocalization with a
coherence length that increases like $N^{1/3}$ in matrix theory
and $N^{1/4}$in the AdS/CFT correspondence.

\section{Holography and matrix theory}

Let us begin with matrix theory. For the present purposes we are
interested in uncompactified matrix theory described by $0+1$
dimensional SYM theory. For a review of matrix theory  and
notations we refer the reader to \cite{8}.

Matrix theory can be thought of as the Discrete Light Cone
Quantization (DLCQ)\cite{9} of M-Theory in which the spacetime is
compactified on an almost light like circle $X^-$. The discrete
conjugate momentum is related to the gauge group rank $N$  by $P_-
R =N$. Thus we see that if we fix the momentum $P_-$, removing the
IR cutoff (letting $R \to \infty$) is tantamount to letting $N \to
\infty$.

What has not been sufficiently realized is that $N$ also plays the
role of an infrared cutoff in the transverse dimensions. To see
why, let us first consider the 10 dimensional metric and dilaton
describing a collection of $N$ coincident D0-branes in the near
horizon limit \cite{10}.
\begin{eqnarray}
ds^2 &=& f^{- 1/2} dt^2 +f^{1/2}dx^i dx^i\cr \exp(2\phi)&=&
f^{3/2}\cr f&=&{Nl_{11}^9 \over R^2 r^7}
\end{eqnarray}
where $l_{11}$ is  the 11 dimensional Planck scale.

Now consider the limits of validity of (2.1). At small $r$ the ten
dimensional supergravity description breaks down because the
effective string coupling gets large. In 11 dimensional terms, the
local value of the radius of the 11th direction becomes bigger
than $l_{11}$. This happens at $r \sim N^{1/7}l_{11}$

While we have to give up the duality between 10D supergravity and
D0-brane physics at this point we can replace it with a duality
between D0-branes and 11D supergravity. This is the basis for
matrix theory. Thus there is no limit on the matrix
theory/Supergravity duality at  $r \sim N^{1/7}l_{11}$

At the large distance end another limitation is reached. The
scalar curvature ${\cal R}$ of the 10D metric satisfies

$${\cal R} \sim {r^{3/2}R \over N^{1/2} l_{11}^{9/2}} $$

It is monotonically increasing with $r$ and exceeds the string
scale at $r \sim N^{1/3}l_{11}$. At this point the supergravity
description completely breaks down. The  region $r> N^{1/3}l_{11}$
is the region where the D0-brane quantum mechanics can be treated
perturbatively. From the supergravity point of view $r=
N^{1/3}l_{11}$ represents an infrared cutoff beyond which
classical supergravity is no longer applicable. This means that
when two colliding objects  in matrix theory approach each other
from infinity,  semiclassical gravity will not generally describe
their  interactions correctly until $r<N^{1/3}l_{11}$. Although it
is true that matrix theory with 16 supersymmetries sometimes
agrees with DLCQ tree graph supergravity to asymptotic distances
this probably has more to do with the tight constraints of maximal
supersymmetry than with any general reason for agreement. A more
typical example is matrix theory on a blown up orbifold where for
finite $N$ the supergravity and one loop matrix theory disagree
\cite{11}. If this case is typical, we would expect agreement only
when $N>({r \over l_{11}})^3$.

These considerations suggest although the compactification radius
$R$ is allowed to be vanishingly small, the D0-branes create a
bubble of space whose transverse size grows as $N^{1/3}$ so that
the limit $N \to \infty $ is effectively decompactified.  It is
therefore interesting to ask if we can see the scale
$N^{1/3}l_{11}$ occurring in matrix theory. In the original matrix
theory conjecture \cite{3} it was speculated that the threshold
bound state describing a supergraviton would grow with $N$. One
estimate was based on the well known $v^4/r^7$ velocity dependent
effective interaction between D0-brane clusters and  suggested
that the bound state radius grows like $N^{1/9}l_{11}$. A second
estimate based on a perturbative large N argument gave the even
more rapid growth $N^{1/3}l_{11}$. Recently Polchinski has given a
rigorous proof \cite{12} that the growth is at least as fast as
$N^{1/3}l_{11}$. Polchinski's  argument is based on the virial
theorem. The argument I will give here is a less rigorous
paraphrase of Polchinski's but gives some intuition about the
nature of the bound state.

Let us use the gauge freedom of matrix theory to work in a basis
in which one of the 9 X-coordinates, say $X_{1}$,  is diagonal.
The eigenvalues can be thought of as the locations  of the
constituent D0-branes along the $X_1$ axis. Let us suppose that
they are smoothly spread over a region of size $L$. Now consider
the quantity $\langle Tr (X_1)^2\rangle$. This  obviously
satisfies
\begin{equation}
\langle Tr(X_1)^2\rangle \sim N L^2
\end{equation}

Consider the quantity $\langle TrY^2 \rangle$ where $Y$ is any of
the other 8 $X's$. The off diagonal elements of the matrix$Y$ are
described by harmonic oscillators in the background of $X$ with
frequency of order
\begin{equation}
\omega \sim {LR \over l_{11}^3}
\end{equation}
and fluctuation $(\Delta Y)^2 \sim {l_{11}^3 \over L}$. Since
there are of  $\sim N^2$ such elements we find
\begin{equation}
\langle TrY^2\rangle \sim {N^2 l_{11}^3 \over L}
\end{equation}
But now we can use rotational symmetry to equate $\langle
Tr(X_1)^2\rangle$ and $\langle TrY^2\rangle $ giving
\begin{equation}
L \sim N^{1/3}l_{11}
\end{equation}

The typical conjugate momentum of a matrix element is also easily
estimated and is given by
\begin{equation}
\Delta P_{ij} = N^{1/6}/l_{11}
\end{equation}
Thus we see that the bound state grows large with $N$, extending
to the boundaries of the region of validity of 10D supergravity.
As seen from eq(2.3) the matrices $X$ have very high frequency
oscillations reminiscent of the high frequency zero point
oscillations of free strings which also lead to a growth of the
wave function but in this case only a logarithmic growth
\cite{13},\cite{14}. Finally, the  kinetic energy of the D0-branes
is estimated as follows. The total kinetic energy is ${R \over 2 }
Tr P_{\perp}^2$. Using (2.6) and the fact that there are $N^2$
matrix elements we find the total kinetic energy to be of order
${R N^{7/3} \over l_{11}^2}$. This is to be compared with the
typical energy scale in DLCQ M-theory ${R \over N {l_s}^2}$.
Evidently, on the scale of the energies of physical processes the
kinetic energies are huge. The kinetic energy per D0-brane is
\begin{equation}
E/N={R N^{4/3} \over l_{11}^2}
\end{equation}
This enormous energy is cancelled by the quartic and fermionic
terms in the hamiltonian but this estimate gives an idea of the
energy scales involved.

In light of the above, let us consider a collision between two
gravitons. Most of the literature on scattering in matrix theory
makes the implicit assumption that the "wave function effects" are
not important. What this means is that the scattering objects are
described by little clusters of D0-branes which are much smaller
than the distance separating them. As we shall see this is
completely incorrect.

For simplicity take the gravitons to have equal light cone momenta
and therefore equal values of $N$. In the transverse center of
mass frame they have equal and opposite transverse momenta
$P_{\perp}$   and $-P_{\perp}$. The light cone energy is
\begin{equation}
E_{lc} ={P_{\perp}^2 \over P_-} =R{P_{\perp}^2 \over N}
\end{equation}
and the Mandelstam invariant center of mass energy is
\begin{equation}
S=2P_{\perp}^2
\end{equation}

Suppose $P_{\perp}$ is fixed and of order $1/l_{11}$. If Matrix
theory is consistent then the scattering amplitude must tend to a
finite limit in 11D Planck units as $N$ increases. But as we have
seen the size of the bound state wave functions grow as $N^{1/3}$.
Each particle is huge blob of eigenvalues and the blobs begin to
overlap long before the particles come close in the usual sense.
During the period of overlap the constituents of each blob lose
their identity. This is obvious because of the very large energy
scales involved in the D0-brane dynamics eq(2.7). The puny
available energy in eq (2.8) is not enough to significantly modify
the correlations in the ground state. Therefore the state of the
system should more closely resemble the ground state of the $2N
\times 2N$ matrix theory than two overlapping but distinguishable
subsystems.

Thus the history of the scattering process has two very different
but equivalent descriptions. In the usual space time supergravity
description two small particles come in from infinity and remain
essentially noninteracting until they come within a distance of
order $l_{11}$. They interact for a short time and then separate
into  final particles which cease to interact as soon as they are
separated by $l_{11}$. In light cone units the interaction lasts
for a time ${l_{11}N \over P_{\perp}R}$.

The holographic matrix description also begins with asymptotically
distant noninteracting objects. In this description the
constituents begin to merge and interact when their separation is
of order $N^{1/3}l_{11}$. As they approach, the many body wave
function begins to more and more resemble the ground state. The
system remains in this entangled state for a light cone time of
order ${l_{11}N{4/3} \over P_{\perp}R}$ and then separate into
noninteracting final clusters. The situation  is particularly
perplexing if the energy is not very large and the impact
parameter is much larger than $l_{11}$. In this case the gravity
description the particles miss each other and just continue
without significant deflection.  Exactly how this miracle happens
from the SYM description is still a mystery.  We will see exactly
the same puzzles in the AdS/CFT correspondence.

\setcounter{equation}{0}
\section{Holography and the AdS/CFT Duality}

String theory in $AdS_5 \times S_5$ is dual to SYM theory on the
boundary of the space \cite{4}, \cite{5}. As pointed out by
Witten, this is another example of a holographic connection
\cite{6}. For our purposes AdS is best thought of as a finite
cavity with reflecting walls. The metric is given by
\begin{equation}
ds^2 = R^2 dS^2
\end{equation}
where $R$ is the radius of curvature of the AdS and $dS^2$ is the
metric of a  "unit" AdS. The unit AdS metric is
\begin{equation}
dS^2 ={(1+r^2)^2 \over (1-r^2)^2}dt^2 - {4\over (1-r^2)^2}(dr^2
+r^2 d\Omega)
\end{equation}
with $d\Omega$ being the unit 3-sphere.

The full geometry is $AdS_5 \times S_5$. The $S_5$ factor is a
5-sphere of radius $R$. Although the boundary, $r=1$ is an
infinite proper distance from any point in the interior of the
ball $r<1$ the time for a light signal to  reflect off the
boundary is finite. A light signal originating at $r=0$ (with
vanishing $S_5$ momentum) will return after a coordinate time $\pi
$. Thus as far as light signals are concerned the space behaves
like a finite radiation cavity with reflecting walls.  The effect
of the inhomogeneous metric is to slow the light velocity at the
center to half its value at the boundary. In other words the bulk
sphere has a varying dielectric constant. A very simple example of
the equivalence of bulk physics with the boundary theory is given
by the restriction of causality. Consider a signal originating at
a point on the boundary. At a later time it will reappear at the
antipodal point on the boundary. In the SYM description it travels
with the speed of light on the boundary taking a time $\pi $ to
get to the antipode. In the dual bulk theory the signal travels
through the center of the ball, $r=0$, along a light like
geodesic. A simple calculation shows that it again arrives after
coordinate time $\pi $.

Massive particle trajectories (timelike geodesics) are all
periodic in time  with period $2 \pi $. These trajectories never
reach the boundary. The cavity walls repel massive particles with
a force which diverges near $r=1$. The force is proportional to
the mass of the particle as is always the case in gravity. From
the point of view of the $AdS_5$ the particles carrying momentum
along the 5-sphere are massive. A massive particle which starts at
$r=0$ with velocity $v$ will move outward on a radial trajectory
for a time $\pi /2$ at which point it reaches a maximum radial
coordinate satisfying
\begin{equation}
v^2 = {4r_{max}^2 \over (1+r_{max}^2)^2}
\end{equation}
In describing the SYM theory we will use the dimensionless metric
$dS^2$. This means that all  SYM quantities will be treated as
dimensionless. The corresponding quantities in the bulk theory
carry their usual dimensions. To go from one to the other the
conversion factor is $R$. For example an SYM energy of order $1$
corresponds to an energy of order $1/R$ in the bulk theory. A
coordinate time interval $t$  is an interval $Rt$ in bulk units.

The dimensionless parameters of the bulk theory are the 10
dimensional string coupling constant $g_s$ and the ratio of the
radius of curvature to the string length scale  $ R/l_s$. The
parameters of the dual SYM theory are the SYM coupling $g_{ym}$
and the rank of the gauge group $N$. The connection between these
parameters was given by Maldacena,
\begin{eqnarray}
g_s &=& g_{ym}^2 \cr R/l_s &=& (Ng_s)^{1/4}
\end{eqnarray}

The fact that by increasing $N$  the radius of curvature in
eq(3.11) can be made to increase while keeping the string coupling
fixed leads to a conjecture for a new nonperturbative definition
of IIB string theory in terms of SYM theory. The $AdS \times S_5$
geometry can be thought of as an infrared regulator for type IIB
string theory. As $R \to \infty$ the space becomes locally flat 10
dimensional Minkowski space. To formulate this precisely let us
begin with Euclidean SYM theory in the Euclidean version of the
metric (3.2).
\begin{equation}
dS^2 ={-(1+r^2)^2 \over (1-r^2)^2}d\tau^2 - {4\over
(1-r^2)^2}(dr^2 +r^2 d\Omega)
\end{equation}
I will refer to these coordinates and their Minkowski counterparts
as ''cavity coordinates" where $\tau$ is Euclidean time. It is
very convenient to transform to ''1/2-plane" coordinates with
metric
\begin{equation}
ds^2 = -R^2 {(dx^i dx^i + dy^2) \over y^2}
\end{equation}
The 4 noncompact coordinates $x^i$ are parallel to the boundary
and can also be used as coordinates for the SYM theory. The
coordinate $y$ runs perpendicular to the boundary and varies from
zero to infinity.

The transformation from 1/2-plane to cavity coordinates is given
as follows. First transform $(x^i,y)$ to 5 dimensional polar
coordinates $\rho,  \theta, \alpha, \beta, \gamma,  $.
\begin{eqnarray}
y&=& \rho \cos \theta \cr x^1&=&\rho \sin \theta \cos \alpha \cr
x^2&=& \rho \sin \theta \sin \alpha \cos \beta \cr x^3&=& \rho\sin
\theta \sin \alpha \sin \beta \cos \gamma \cr x^4&=&\rho \sin
\theta \sin \alpha \sin \beta \sin \gamma
\end{eqnarray}
Now set $\rho = e^{\tau}$ and $\cos \theta = {1-r^2 \over 1+r^2}$.
The three angles $\alpha, \beta, \gamma$ are the coordinates of
the unit sphere $\Omega$.

We will be interested in correlation functions of various fields
in the superconformally invariant SYM theory. Thus consider a set
of points $x_a$ on the boundary of the 1/2-plane coordinates. For
each pair of points $a,b$ define  $x_{ab} \equiv  |x_a - x_b|^2$.
In terms of Euclidean cavity coordinates $x_{ab}$ is given by
\begin{equation}
x_{ab} = e^{(\tau_a + \tau_b)}(\cosh\tau_{ab} - cos\phi_{ab})
\end{equation}
where $\tau_{ab} \equiv \tau_a -\tau_b$ and $\phi_{ab}$ is the
angular separation between the points in $\Omega$. It is also
convenient to define $Z_{ab} =(\cosh\tau_{ab} - cos\phi_{ab}) $.

Euclidean correlation functions of the SYM theory are typically
homogeneous functions of the $x_{ab}$  of degree determined by the
dimensions of the operators. To express the corresponding
correlators in cavity coordinates  just replace each $x_{ab}$ by
$Z_{ab}$. The various of $e^{\tau}$ cancel the Jacobian factors in
the transformation of fields with nonvanishing dimensions. Thus
the correlators are homogeneous functions of the $Z_{ab}$. As an
example, the correlation function of two scalar fields $\Phi$ of
dimension 4 is of the form  $Z_{ab}^{-4}$

It is now a simple matter to pass to Minkowski signature by
replacing $\tau $ by $it$. Thus the correlator becomes
\begin{equation}
\langle \Phi(x_a) \Phi(x_b) \rangle = (\cos t_{ab} -
cos\phi_{ab})^{-4}
\end{equation}
The singularity when $\cos t_{ab} = cos\phi_{ab}$ is the usual
light cone singularity.

Strictly speaking there is no true S matrix in AdS space. As I
have emphasized, AdS is for all practical purposes a finite cavity
with reflecting walls. Asymptotic states can not be defined in
such a geometry. The strategy that we follow is to introduce
sources on the walls of the cavity which act as particle sources
and detectors. This will allow us to define a finite time version
of the S matrix. When the size of the box is allowed to increase,
keeping fixed the energies, impact parameters and other physical
quantities the finite time S matrix should tend to a true
asymptotic scattering amplitude.

Before discussing the boundary sources further we need to
determine what quantities should be kept fixed as $R \to \infty$
in order to recover flat space string theory.  First of all we
must keep the microscopic parameters of string theory fixed. This
means letting $R/l_s \to \infty$ with $g_s$ fixed. In terms of
Yang Mills quantities
\begin{eqnarray}
g_{ym}&=& fixed \cr N &\to& \infty
\end{eqnarray}
In addition the energy scale of physical processes should be fixed
in string units. In terms of the dimensionless energy of the SYM
theory $E$
\begin{equation}
E\sim (g_{ym}^2 N)^{1/4}
\end{equation}
Thus we see that the flat space limit involves the high energy
limit of large $N$ SYM theory. We will also require restrictions
on the angular momenta of particles.

We will define a spacetime region called the ''lab". The lab is
centered at $r=t=0$. Its linear dimensions $L$ in both space and
time are fixed in string units but are much larger than $l_s$ At
the end we may take $L/l_s$ as big as we like. As $N \to \infty$
the entire region of the lab becomes accurately described by flat
spacetime. The sources will be constructed in such a way as to
insure that the entire collision process takes place within the
lab.

A particle can carry momentum components in both the $AdS_5$
directions and in the $S_5$ directions. We will call these $p$ and
$k$ respectively. For the moment we will ignore $k$. Consider a
massless particle that is inside the lab with momentum $p$. Its
angular momentum $l$ is necessarily less than $Lp$. Since the
cavity is spherically symmetric, The angular momentum of a freely
moving incoming particle is conserved. Therefore if the particle
is to arrive in the lab it must be emitted from the boundary with
$l<Lp$. This restriction guarantees that the "beam" is focused to
pass through the lab.

As an example we will consider scattering amplitudes for dilatons
carrying vanishing momentum in the $S_5$ directions.  The dilatons
will be emitted in such a way that they propagate freely toward
the region $r=0$ where they meet and interact within the lab.
Since the wavelength of the particles is vanishingly small by
comparison with the radius of the AdS space, the propagation of
the wave packets toward the lab can be treated by geometrical
optics. The time it takes for a wave packet to travel from the
boundary to the lab is $\pi/2$, just half the time for a light
signal to cross the AdS space. Therefore the initial sources must
act at $t=-({\pi \over 2} \pm {L \over R})$. Similarly the final
detector-sources must act at  $t=+({\pi \over 2} \pm {L \over
R})$.

The appropriate SYM operators for emitting all massless 10
dimensional particles are known. In particular the operator that
creates a dilaton at the boundary is the dimension 4 operator $Tr
F_{\mu \nu } F^{\mu \nu}\equiv FF$. Let us consider the emission
operator for a zero angular momentum dilaton of  bulk energy $p$.
The obvious choice is
\begin{equation}
A^{in}(p) \sim \int dt d\Omega e^{ipRt}FF
\end{equation}
However in order to build wave packets which arrive at the lab at
$t=0 \pm {L \over R}$ we need modify the definition of $A$. This
can be done by replacing the factor $e^{ipRt}$ by a wave packet of
finite extent. Let $f_{in}\left[(t-{\pi \over 2}){R \over L}
\right]$ be a smooth function (such as a gaussian) which is peaked
at $t=\pi/2$. The definition of $A $ is
\begin{equation}
A^{in}(p) \sim \int dt d\Omega  f_{in}\left[(t-{\pi \over 2}){R
\over L} \right] e^{ipRt}FF
\end{equation}
A similar expression defines the operators representing the final
particles.
\begin{equation}
A^{out}(p) \sim \int dt d\Omega  f_{out}\left[(t+{\pi \over 2}){R
\over L} \right] e^{-ipRt}FF
\end{equation}
To create particles of arbitrary angular momentum the integral
over $\Omega$ should contain the relevant $O(4)$ spherical
harmonic.

The recipe for computing bulk S matrix elements from SYM
quantities is straightforward.
\begin{equation}
S=\langle 0|\prod_{out}Z_{out} A^{out} \prod_{in} Z_{in} A^{in}
|0\rangle
\end{equation}
The factors $Z$ are inverse boundary-bulk  propagators which  are
needed to amputate the external AdS propagators.

The above prescription for recovering flat space amplitudes can be
generalized to include nonvanishing momenta along the 5-sphere.
The operators which create particles with nonvanishing $O(6)$
angular momentum $n$ are schematically of the form
$
Tr FF XXXX...
$
where $F$ represents components of the Yang Mills Field strength
and $XXXX..$ is a polynomial of order $n$ in the scalar fields
which transform as vectors under $O(6)$. These are operators of
mass dimension $4+n$. We must also integrate these operators with
functions of time and $\Omega$ in order to project out definite
energy and $O(4)$ angular momentum. Again the frequencies should
be of order $g_{ym}^2 N^{1/4}$ in order to keep the physical
momentum of the bulk particles of order unity in string units.

Thus we see that passing to the flat space limit generally
involves operators in the SYM  theory which are high frequency
components of high dimension operators.

To actually compute scattering amplitudes from conformal field
theory data, a useful strategy might be to use the operator
product expansion for the operators $A^{in,out}$. Consider for
example a two particle scattering process in which the incoming
(outgoing) particles are emitted (absorbed)  at time $t_{in,out}
=\pm\pi/2$. The angular positions of the incoming particles are
$\Omega_{1,2}$ and the outgoing particles $\Omega_{3,4}$. The 4
points $(1,2,3,4)$ are far from each other in spacetime and it is
not obvious why the operator product expansion is useful. However,
consider the case where there is a small momentum transfer
$(p_1-p_3)<<p$. Then the locations of $1$ and $3$ will be almost
light-like with respect to each other. In the rules for
continuation from Euclidean to Minkowski signature in AdS space
the almost light like separation between $1$ and $3$ maps to an
almost vanishing Euclidean separation so that the OPE should
provide an expansion for small angle scattering. Obviously, the
operators of low dimensionality in the operator product of
$A(1)A(3)$ correspond to  massless exchange. In addition we also
expect contributions corresponding to massive string exchange with
masses of order $l_s^{-1}$. From the point of view of the operator
product expansion this means  operators of dimensionality $\sim
g_s N^{1/4}$. We will leave it to a future publication, hopefully
by someone else, to work out the detailed rules for computing on
shell scattering amplitudes from CFT data in the flat limit.

\setcounter{equation}{0}
\section{The Infrared Ultraviolet Connection}
The connection between the boundary SYM theory and the ideas of
holography rely on an important connection between the ultraviolet
behavior of the SYM theory and the infrared behaviour of the bulk
supergravity \cite{7}. We will begin by reviewing the argument for
counting the number of degrees of freedom of the system. By now it
is well known that an ultraviolet cutoff at wavelength $\delta$ in
the SYM is equivalent to a cutoff in the radial coordinate r at
$r=1-\delta$. The cutoff SYM describes the bulk supergravity in
the interior of the ball $r< 1-\delta$. Now the number of cutoff
cells of coordinate size $\delta$ on the boundary of this ball is
of order $1/{\delta}^3$. Assuming that each independent SYM field
has one degree of freedom per cells, the total number of degrees
of freedom is
\begin{equation}
N_{dof} \sim {N^2 \over \delta^{3}}
\end{equation}
If we use
\begin{eqnarray}
R&=&l_s (Ng_s)^{1/4} \cr Area &=& {R^8 \over \delta^3} \cr G&=&
g_s^2 l_s^8
\end{eqnarray}
where $Area$ is the area of the cutoff 3-sphere times $S_5$ and $G
$ is the 10D gravitational coupling constant we find the typical
holographic behaviour
\begin{equation}
N_{dof} \sim {Area /G}
\end{equation}

Let us push this reasoning to the extreme and take the cutoff
$\delta$ such  that the proper volume of the  4 dimensional ball
$r<\delta$ is $R^4$ (dimensionless volume $\sim 1$). The area of
the cutoff boundary sphere is then $\sim R^3$ and the number of
degrees of freedom is just $N^2$. In other words it takes $N^2$
degrees of freedom to describe all the states of the bulk theory
which are supported within a sphere of proper size $R$. This means
that the physics, within a neighborhood small enough so that
curvature can be ignored, is coded by the $N^2$ matrix degrees of
freedom and that the inhomogeneous spatial modes of the SYM are
unexcited. This suggests the possibility that the states supported
within such a neighborhood might be described by an $N\times N$
matrix quantum mechanics.

As an illustration of the IR-UV connection consider a graviton
carrying momentum $k$ along the $S_5$ and momentum $p>>k$ in the
radial direction along $r$. Its total energy is $E = \sqrt{p^2
+k^2}$. It is created by applying the operator
\begin{equation}
A^{in}(p,k) = \int dt d\Omega  f_{in}\left[(t-{\pi \over 2}){R
\over L} \right] e^{iERt} Tr FFXXXXX...
\end{equation}
where there are  $n=kR$ factors inside the trace. This means that
the total energy is divided among $n$ SYM quanta and the energy
$\nu$ of each SYM quantum is
\begin{equation}
\nu = \sqrt {(p^2 + k^2) \over k^2} \approx {p \over k}
\end{equation}
This corresponds to a cutoff in the SYM theory $\delta \approx
1/\nu \approx {k \over p}$. The implication is that in the bulk
theory the particle created by $A$ appears not at the boundary but
at $1-r \approx {k \over p}$. This makes good sense for the
following reason. From the point of view of $AdS_5$ a particle
with $S_5$ momentum $k$ is a massive particle with mass $k$. The
classical trajectory of such a massive particle with (bulk) energy
$E$ has a turning point (vanishing velocity) given by (3.3). In
terms momentum components the turning point is at $1-r \approx {k
\over p }$. Hence the particle starts out at the outermost point
on its trajectory.

The fact that massive particles originate in the interior of the
AdS space does not require a modification of the rules for
constructing the $A$ operators. Although they start closer to the
interaction point at $r=0$, the time that it takes to arrive at
$r=0$ is independent of the mass.

From the above discussion it seems that the cutoff theory with a
given value of $\delta$ can describe the sector of the theory
containing particles with ${k \over p}\geq \delta$. Suppose for
example, all the particles in a given reaction have ${k \over p}
\sim 1$.In this case only the lowest modes of the SYM theory are
excited corresponding to configurations which are spatially
homogeneous (in the boundary theory). In other words physical
processes involving such particles always appear completely
smeared  and nonlocal in the holographic SYM description. The
situation is very similar to the matrix case.

At this point it is interesting to consider just what problems we
in principle would know how to set up and solve if we could
completely master SYM theory and find all its correlators, and
energy levels. First of all we could apply the recipe described in
section(3) to compute any scattering amplitude involving 10
dimensional massless particles. Since we do not expect any other
stable particles in the theory, this exhausts all IIB flat space
scattering amplitudes.

In addition we could compute the thermodynamics of the theory and
discover the existence of a phase transition at (dimensionless)
temperature $\sim 1$. This corresponds to the formation of a large
black hole at bulk temperature $\sim {R^{-1}}$. This object has no
significance in the flat space limit since it has a size of order
the radius of curvature.

However most of ordinary flat space physics would remain out of
reach even though it implicitly must be described by the SYM
theory. As an example consider the problem of describing an
ordinary 10 dimensional Schwarzschild  black hole of finite mass
and entropy as $N \to \infty$. For simplicity the black hole could
be located near the center of the AdS at some point on the
5-sphere. If the proper distance of the black hole from the center
is kept fixed as $R \to \infty$ then its coordinates will tend to
the origin at $r=0$. The image will become completely symmetric on
the 3-sphere. What features of the SYM state contain the
information of the exact position or even the fact that it is a
black hole or any other object of the same mass and angular
momentum is not known. In fact it is not even clear how to
distinguish this configuration from a pair of distant black holes
or other objects of the same total mass if their separation is
much smaller than the radius of curvature $R$. In principle the
AdS/CFT correspondence requires the SYM theory to contain these
objects. Recognizing them from their SYM description requires
deciphering the holographic code.

\setcounter{equation}{0}
\section{Instantons in AdS}

It has been suggested that a  simple place to begin trying to
crack the holographic code is the theory of instantons. In the
bulk supergravity theory the instantons are D-instantons whose
holographic images are expected to be ordinary Yang Mills
instantons . In order to discuss instantons we must consider the
Euclidean version of AdS. As is well known this is a 5 dimensional
ball bounded by a 4-sphere of radius $R$. The Euclidean SYM theory
lives on this boundary sphere. As before, when discussing the SYM
the 4-sphere will be thought of as having unit radius.

For small $\delta$, a D-instanton in the bulk located at
$r=(1-\delta)$ is represented in the SYM as an localized instanton
of size $\rho = \delta$ \cite{15}. This is another example of the
UV-IR connection.  D-instantons near the center of the ball
correspond to Y.M. instantons which fill the entire boundary
sphere homogeneously. In fact the gauge field for such instantons
is completely homogeneous up to a gauge transformation. Note that
if we have one or more D-instantons at a fixed separation from one
another and from $r=0$ then as $N$ increases their coordinates get
closer and closer to $r=0$. Thus in the limit all the instantons
lying within a finite proper volume are found within an
infinitesimal coordinate distance from the origin. Thus in the SYM
theory they are described by the largest homogeneous gauge
instantons. Again, the features of the SYM description which
distinguish the different D-instanton configurations are obscure.

An SYM instanton will appear approximately homogeneous if the
corresponding D-instanton is within a ball of volume $R^5$. An
interesting paradox occurs if we ask how many D-instanton's can we
put in such a region. The naive answer if that the number should
roughly be the 10 dimensional volume of the product of  the
ball$\times S_5$. In other words the maximum instanton number in
this volume is naively $({R \over l_s})^{10} = (N g_s)^{5/2}$.
However if we try to build homogeneous instanton configurations on
the SYM sphere we find that we can have a maximum instanton number
of order $N$. A single large $SU(2)$ instanton is a homogeneous
configuration. If we try to put two instantons into the same
$SU(2)$ subgroup we find that it can not be done without making
the field inhomogeneous. Since their are $N/2$ commuting $SU(2)$
subgroups we can only accommodate this number of homogeneous
instantons. Either the D-instantons within volume $R^5$ can not be
identified with homogeneous gauge field configurations or there
must be some reason why it is not possible to cram as many
D-instantons into a region as the naive argument suggests.

Exactly  this conclusion can be reached by an argument similar to
that used for D0-branes in section 2. If we are interested in a
small region of AdS over which the curvature can be ignored we can
use a flat space description. D-instantons are formally D-branes
of dimensionality $-1$ and are described by a matrix integral
\cite{16} defined by the dimensional reduction of maximally
supersymmetric SYM. If the instanton number is $k$ the matrices
are $k$ by $k$. The action for the D-instantons is
\begin{equation}
S=-{1 \over {g_s l_s^4}}Tr[X^i,X^j]^2
\end{equation}
An argument exactly paralleling the one in eqs(2.2), (2.3) and
(2.4) will give  the size of the region occupied by $k$
D-instanton's. As in that case, first diagonalize $X^1$ and assume
that the eigenvalues are smeared over a region of size $L$. As
before
\begin{equation}
\langle Tr(X^1)^2\rangle \sim k L^2
\end{equation}

Now consider one of the off diagonal elements of $Y$ where $Y$ is
any other $X^j$. The average of  $Y^2$ in the  $X^1$ background is
\begin{equation}
\langle Y^2\rangle =g_s l_s^4 /L^2
\end{equation}
and
\begin{equation}
\langle Tr Y^2 \rangle =k^2 g_s l_s^4 /L^2
\end{equation}

Using rotational symmetry to equate (5.23) and (5.25) gives
\begin{equation}
L=(kg_s)^{1/4} l_s
\end{equation}
To find the maximum number of D-instanton's that we can put into a
flat region of size $R $ we set $L=R= (Ng_s)^{1/4}$. Thus we find
that the maximum number of D-instanton's is of order $N$ in
agreement with the maximum number of homogeneous gauge field
instantons. It is also clear that the positions of the
D-instanton's, within the region approximated by flat space, in
the limit $N\to \infty$ must be coded somehow in the $N \times N$
matrix degrees of freedom of the SYM and not in the large gauge
field inhomogeneities.

Finally, classical gravitational considerations give the same
result. The gravitational field of $k$ D-instantons  in flat space
is given by
\begin{equation}
ds^2= f^{1/2}dx^i dx^i
\end{equation}
where
\begin{equation}
f=1+kg_sl_s^8 r^{-8}
\end{equation}
Thus the gravitational field extends out to distance
$r=(kg_s)^{1/4} l_s$. If this distance is not to exceed $R$ then
$k\leq N$.

\section{Decoding the Hologram}

Exactly how information  is holographically stored in either
matrix theory or the AdS/CFT correspondence is a mystery. I will
try to give some thoughts about it. Lets begin with matrix theory.

The  $N^{1/3}$ increase in the size of the low energy wave
functions of D0-brane wave functions is caused by the ground state
oscillations of an increasing  number ($\sim N^2$) of high
frequency modes. The situation is parallel to that in free string
theory where as the number of modes is increased the ground state
expands \cite{13}. In the free string case the expansion is only
logarithmic but for any finite coupling it will eventually grow
like a power \cite{2}. The increase in the sizes of images
eventually blurs the details of a system. For example if the
system consists of two distinct objects separated by transverse
distance $\Delta$ then when $N^{1/3} > {\Delta \over l{11} }$ the
holographic images become entangled. Given a state of the matrix
theory at very large $N$ it would be very difficult to decipher
its meaning.

The trick in decoding the hologram is to get rid of the high
frequency oscillations. This can be done by averaging over time
but the right thing has to be averaged. For example we could
define a density of D0-branes along the $X^1$ axis in terms of the
distribution of its eigenvalues. This however is a very slowly
varying quantity which does not have high frequency oscillations.
The right thing to average is the Heisenberg operators
representing the matrix elements $X_{a,b}$. For example,  we  may
average $X$ over a time $\delta t$. If we work in the eigenbasis
of the Hamiltonian, the averaging is equivalent to throwing away
all (quantum) matrix elements $\langle E_1|X_{ab}|E_2\rangle$ with
$|E_1 -E_2| > 1/{\delta t}$. The resulting quantum operators will
have a modified distribution of eigenvalues. Since modes of
frequency $> 1/{\delta t}$ are now absent the distribution should
have a smaller spread. Therefore the holographic image of several
objects should become clear. It is evident that all of this is a
manifestation of the stringy space, time uncertainty relation
\cite{17}.

In order to be a little more quantitative I will make an
assumption that is motivated by a particular view of the large $N$
limit. According to this view, the large $N$ limit is a fixed
point of a kind of renormalization group associated with
integrating out rows and columns of the $N \times N$ matrix
degrees of freedom to produce a theory with smaller matrices. What
I will assume is that time averaging over $\delta t$, or
equivalently, integrating out high frequency modes is equivalent
to replacing the original $N \times N$ matrix system by another
with smaller $n \times n$ matrices. The maximum relevant frequency
for the original system is given by eq(2.3) with $L =
N^{1/3}l_{11}$. We will call this the characteristic frequency
$\omega_N$.
\begin{equation}
\omega_N ={ N^{1/3}R \over l{11}^2}
\end{equation}
If we identify  the characteristic frequency of the $n \times n$
model to be $\omega_n = {(\delta t)}^{-1}$ then
\begin{equation}
{n \over N}= {\omega_n^3 \over \omega_N^3}= {l_{11}^6 \over (R
\delta t)^3 N }
\end{equation}
Furthermore since the size of the eigenvalue distributions of $X$
scale like $N^{1/3}$ we should find the spread diminished by the
factor ${\omega_n \over \omega_N}$. According to this estimate, by
averaging over $\delta t = l_{11}^2 /R$  resolution of order the
Planck length should be restored for a pair of gravitons.

For the AdS/CFT correspondence decoding the hologram seems to be
very different. In the flat space limit the SYM dimensionless
energy of a given system increases like $(g_sN)^{1/4}$. On the
other hand the information is coded in the longest wavelength
modes on the unit sphere. These modes have frequency $\sim 1$ in
dimensionless units which corresponds to very long bulk time
scales of order $R$. In other words, information seems to be coded
in extremely slow degrees of freedom. At the moment I have no idea
how this works.

\section{Acknowledgements}
Much of what is written in this paper was stimulated during
discussion with Juan Maldacena and Andy Strominger while I was
visiting Harvard in September of this year. This of course does
not mean that they are responsible for the confusion and fuzzyness
of my ideas. I also benefited from many conversations with Steve
Shenker over the last couple of years about these issues.

%%%%%%%%%%%%%%%%%%%%%%%%%%%%%%%%%%%%%%%%%%%%%%%%%%%%%%%%%%%%%%
%%%%%%%%%%%%%%%%%%%%%%%%
%
%     \begin{equation}    \end{equation}
%
%%%%%%%%%%%%%%%%%%%%%%%%%%
%     \begin{eqnarray}
%
%     \end{eqnarray}
%%%%%%%%%%%%%%%%%%%%%%%%%
%%%%%%%%%%%%%%%%%%%%%%%%%%%%%%%%%%%%%%%%%%%%%%%%%%%%%%%%%%%
%\newpage
%

%
%
%
\setcounter{equation}{0}
%\newpage
%

%
%
%
%
%

\begin{thebibliography}{99}
\bibitem{1} G. 't Hooft, Dimensional Reduction in Quantum Gravity,
gr-qc/9310026
\bibitem{2} L. Susskind,  The World as a Hologram, hep-th/9409089
\bibitem{3} T. Banks, W. Fischler, S.H. Shenker, L. Susskind,
M Theory As A matrixModel: A Conjecture, hep-th/9610043
\bibitem{4} Juan M. Maldacena, The Large N Limit of Superconformal
Field Theories and Supergravity, hep-th/9711200
\bibitem{5} S.S. Gubser, I.R. Klebanov, A.M. Polyakov, Gauge Theory
Correlators from Non-Critical String Theory, hep-th/9802109
\bibitem{6} Edward Witten, Anti De Sitter Space And Holography, hep-th/9802
\bibitem{7}  L. Susskind, Edward Witten, The Holographic Bound in Anti-de
Sitter Space hep-th/9805114
\bibitem{8} Daniela Bigatti, Leonard Susskind, Review of Matrix
Theory, hep-th/9712072
\bibitem{9} Leonard Susskind, Another Conjecture about M(atrix) Theory
hep-th/9704080
\bibitem{10} G. Horowitz and A. Strominger, Nucl. Phys. B360 (1990)197.
\bibitem{11} Michael R. Douglas, Hirosi Ooguri, Stephen H.
Shenker, Issues in M(atrix) Theory Compactification,
hep-th/9702203
\bibitem{12} J. Polchinski, unpublished
\bibitem{13} M. Karliner, I. Klebanov and L. Susskind,
Size And Shape Of Strings, Int.J.Mod.Phys. A3 (1988) 1981
\bibitem{14} Leonard Susskind, Particle Growth and BPS Saturated
States, hep-th/9511116
\bibitem{15} Massimo Bianchi, Michael B. Green, Stefano Kovacs, Giancarlo
Rossi Instantons in supersymmetric Yang-Mills and D-instantons in
IIB superstring theory hep-th/9807033 [abs, src, ps, other]
\bibitem{16} M.B. Green ,  Configurations of two D-instantons, Phys.Lett.
B398 (1997) 69
\bibitem{17} Miao LI, Tamiaki Yoneya, D-Particle Dynamics and The
Space-Time Uncertainty Relation hep-th/9611072,  Journal-ref:
Phys.Rev.Lett. 78 (1997) 1219-1222





%
%
\end{thebibliography}
\end{document}